\documentclass[aip,reprint,prl]{revtex4-1}

\usepackage{graphicx}
\usepackage{amsmath}
\usepackage{esint}
\usepackage{tablefootnote}
\usepackage{threeparttable}

\usepackage{natbib}
\usepackage{aas_macros}

\draft 

\begin{document} 

\title{Gravitational orbits in the expanding universe revisited}
\author{V\'aclav Vavry\v cuk}
\affiliation{The Czech Academy of Sciences, Bo\v cn\' i II 1401, 141 00 Praha 4}

\date{\today}

\begin{abstract}
Modified Newtonian equations for gravitational orbits in the expanding universe indicate that local gravitationally bounded systems like galaxies and planetary systems are unaffected by the expansion of the Universe. This result is derived under the assumption of the space expansion described by the standard FLRW metric. In this paper, an alternative metric is applied and the modified Newtonian equations are derived for the space expansion described by the conformal FLRW metric. As shown by Vavry\v cuk (Front. Phys. 2022), this metric is advantageous, because it properly predicts the cosmic time dilation and fits the SNe Ia luminosity observations with no need to introduce dark energy. Surprisingly, the Newtonian equations based on the conformal FLRW metric behave quite differently than those based on the standard FLRW metric. In contrast to the common opinion that local systems resist the space expansion, the results for the conformal metric indicate that all local systems expand according to the Hubble flow. The evolution of the local systems with cosmic time is exemplified on numerical modelling of spiral galaxies. The size of the spiral galaxies grows consistently with observations and a typical spiral pattern is well reproduced.  The theory predicts flat rotation curves without an assumption of dark matter surrounding the galaxy. The theory resolves challenges to the $\Lambda$CDM model such as the problem of faint satellite galaxies, baryonic Tully-Fisher relation or the radial acceleration relation. Furthermore, puzzles in the solar system are successfully explained such as the Pioneer anomaly, the Faint young Sun paradox, the Moon's and Titan's orbit anomalies or the presence of rivers on ancient Mars.
\end{abstract}

\pacs{universe expansion --
      dark energy --
      dark matter --
      FLRW metric --      
      galaxies --
      solar system 
          }

\maketitle 
         
\section{Introduction}

Observations of the cosmological redshift interpreted by Lemaitre \citep{Lemaitre1927} and Hubble \citep{Hubble1929} as an effect of the expansion of the Universe started a new era of cosmology and opened space for applying theory of general relativity (GR) to cosmological problems. Subsequently, the Friedmann equations \citep{Friedmann1922} became the basic equations describing the expanding history of the Universe. Immediately, cosmology was faced with the following fundamental questions: How does the global expansion of the Universe affect local gravitational systems? How do local gravitational fields interact with the expansion and where is a size threshold between systems affected by and resisting the global expansion? Do galaxies and galaxy clusters expand or not? How does the global expansion affect our solar system? These theoretical problems paid attention of many cosmologists, because they have essential consequences for understanding the Universe evolution and for interpreting cosmological observations \citep{McVittie1933, Einstein_Straus_1945, Dicke_Peebles_1964, Noerdlinger_Petrosian_1971, Carrera_Giulini_2010}.  

The simplest problem is to study the Newtonian equations of motion for two-point particles placed in the expanding space and mutually attracted by the gravitational force. If the gravitational field is weak and velocities of particles are non-relativistic, the problem can be solved by perturbations \citep{McVittie1933, Noerdlinger_Petrosian_1971, Bolen2001, Faraoni_Jacques_2007}. In this case, we assume that the metric of the space expansion is perturbed by a weak gravitational field. Assuming that the space expansion is described by the standard Friedmann-Lemaitre-Robertson-Walker (FLRW) metric, the metric tensor $g^{\mu\nu}$ of a gravitational field produced by a point mass $M$ situated in the expanding space reads (see eq. 11 of Noerdlinger and Petrosian \citep{Noerdlinger_Petrosian_1971}) 
\begin{equation}\label{eq1}
\begin{gathered}
ds^2 = -c^2(1+2\alpha)dt^2 + \left((1-2\alpha)\frac{dr^2}{1-kr^2}+r^2d\Omega^2\right)\,,\\ 
 d\Omega^2 = d\theta^2+\sin^2\theta \, d\phi^2 \,,
\end{gathered}
\end{equation}
where $t$ is time, $c$ is the speed of light, $k$ is the Gaussian curvature of the space, $r$ is distance, and $\theta$ and $\phi$ are the spherical angles. Parameter
\begin{equation}\label{eq2}
\alpha = -\frac{G M}{r c^2}\,, \, |\alpha| \ll 1\,,
\end{equation}
is the Newtonian gravitational potential normalized to $c^2$, and $G$ is the gravitational constant. Assuming a massive nonrelativistic particle ($v \ll c$) orbiting in the gravitational field and using the geodesic equation, we finally obtain (see eqs 12ab of Carrera and Giulini \citep{Carrera_Giulini_2010})
\begin{equation}\label{eq3}
\ddot{R} = -\frac{G M}{R^2} + \frac{L^2}{R^3} + \frac{\ddot{a}}{a}R\,,
\end{equation}
\begin{equation}\label{eq4}
L = \mathrm{const}\,,
\end{equation}
where $L = R V^\theta$ is the proper angular momentum, and $V^\theta$ is the proper tangential velocity. For a more detailed derivation, see Appendix A.

Eqs (3) and (4) are called the modified (or improved) Newtonian equations and they differ from the standard Newtonian equations describing the Kepler orbits by term $\frac{\ddot{a}}{a}R$ in Eq. (3) related to the space expansion. The analysis of the modified Newtonian equations applied to the galaxy dynamics shows that the expansion term $\frac{\ddot{a}}{a}R$ affects the orbits within galaxies negligibly \citep{Faraoni_Jacques_2007}. This result led to the conclusion that galaxies and all smaller gravitational systems are not affected by the space expansion and behave as in the static universe \citep{Dicke_Peebles_1964, Noerdlinger_Petrosian_1971, Cooperstock1998, Faraoni_Jacques_2007, Iorio2013}.

The key in deriving the modified Newtonian equations is the assumption that the space expansion is described by the standard FLRW metric. However, Vavry\v cuk \citep{Vavrycuk_Frontiers_2022} shows this metric is incorrect and inconsistent with observations of the cosmological redshift and cosmic time dilation. He claims that the standard FLRW metric must be substituted by the so-called conformal FLRW metric. To describe the expansion of the Universe properly, the proper time must be replaced by the conformal time in the metric \citep{Endean1994, Endean1997, Ibison2007, Gron_Johannesen_2011a}. Consequently, time is not invariant in the modified metric, but its rate varies during the evolution of the Universe. Without changing the time rate, the frequency of photons propagating in the Universe cannot change during the expansion and the photons cannot be redshifted. This result is convincingly supported by observations of Type Ia supernovae (SNe Ia). If the conformal FLRW metric is applied, the cosmological model fits the SNe Ia luminosity observations with no need to introduce dark energy and to assume an accelerated expansion of the Universe \citep{Vavrycuk_Frontiers_2022}. Considering the conformal time in the FLRW metric has also other important consequences. For example, the cosmological and gravitational redshifts are calculated from the metric tensor by the same formula. Obviously, we can ask a question, whether does a local gravitational field in the conformal FLRW metric behave differently from that in the standard FLRW metric or not.

The aim of this paper is to study local gravitational systems in the expanding space described by the conformal FLRW metric. The gravitational field is assumed weak and velocities of particles are non-relativistic. The improved Newtonian equations are derived using the geodesic equation. It is shown that the local gravitational systems in the conformal FLRW metric behave quite differently than in the standard FLRW metric. In contrast to the common opinion that local systems resist the space expansion, the results show that all local systems expand according to the Hubble flow. The evolution of the local systems is exemplified on numerical modelling of spiral galaxies. The theory predicts flat rotation curves and the observed morphology of spirals. Also, other observations supporting the presented theory are discussed.

\section{Theory}

\subsection{Expanding universe described by the conformal FLRW metric}

Let us assume an expanding universe described by the conformal FLRW metric \citep{Gron_Johannesen_2011a, Vavrycuk_Frontiers_2022}:
\begin{equation}\label{eq5}
\begin{gathered}
ds^2 = a^2(t) \left(-c^2 dt^2 + \frac{dr^2}{1-kr^2}+r^2d\Omega^2\right) \,, \\ 
 d\Omega^2 = d\theta^2+\sin^2\theta \, d\phi^2 \,,
\end{gathered}
\end{equation}
where $a(t)$ is the scale factor defining the universe expansion, $t$ is the conformal (comoving) time, $c$ is the speed of light, $k$ is the Gaussian curvature of the space, $r$ is the comoving distance, and $\theta$ and $\phi$ are the spherical angles. 

This metric is conformal to the static universe and it differs from the standard FLRW metric by assuming not only the space expansion but also time dilation during the universe evolution. This metric is convenient, because it leaves the Maxwell equations unchanged from their form in the Minkowski spacetime \citep{Infeld_Schild_1945, Infeld_Schild_1946, Ibison2007}. Moreover, Vavry\v cuk \citep{Vavrycuk_Frontiers_2022} showed that this metric is capable to describe satisfactorily the universe evolution and fits the Type Ia supernovae observations with no need to introduce dark energy.

Using Eq. (5), the equation of the null geodesics, which describes propagation of photons, $ds^2 = 0$, reads
\begin{equation}\label{eq6}
\begin{gathered}
a^2(t) (-c^2 dt^2 + dl^2) = 0 \,, \\ 
dl^2 =  \left(\frac{dr^2}{1-kr^2}+r^2d\Omega^2\right) \,.
\end{gathered}
\end{equation}

Consequently, we get for the comoving velocity $v$ and the proper velocity $V$ of photons
\begin{equation}\label{eq7}
v = \frac{dl}{dt} = c \,, \,\, V = a c \,.
\end{equation}
The propagation velocity of massive particles is described by the geodesic equation 
\begin{equation}\label{eq8}
\frac{d^2 x^\mu}{ds^2} + \Gamma^{\mu}_{\alpha \beta} \frac{dx^\alpha}{ds} \frac{dx^\beta}{ds} = 0 \,.
\end{equation}
Substituting the distance element $ds$ by the time element $dt =dx^0$, we obtain
\begin{equation}\label{eq9}
\frac{d^2 x^\mu}{dt^2} = -\Gamma^{\mu}_{\alpha \beta} \frac{dx^\alpha}{dt} \frac{dx^\beta}{dt} + \Gamma^0_{\alpha \beta} \frac{dx^\alpha}{dt} \frac{dx^\beta}{dt} \frac{dx^\mu}{dt} \,.
\end{equation}
Considering the metric tensor $g^{\mu \nu}$ needed for calculating the Christoffel symbols $\Gamma_{\alpha \beta}^\mu$ in Eq. (9) defined by Eq. (5), we get
\begin{equation}\label{eq10}
a c^2 \dot{v} + \dot{a} c^2 v - \dot{a} v^3 = 0 \,,
\end{equation}
hence
\begin{equation}\label{eq11}
\frac{\dot{v}}{v (1- v^2/c^2)} = -\frac{\dot{a}}{a} \,.
\end{equation}
Consequently, for a massive nonrelativistic particle ($v \ll c$) we write 
\begin{equation}\label{eq12}
\frac{\dot{v}}{v} = -\frac{\dot{a}}{a} 
\end{equation}
and the comoving velocity $v$ and the proper velocity $V$ read
\begin{equation}\label{eq13}
v = v_0 a^{-1} \,, \, \, V = a v = V_0 \,,
\end{equation}
where subscript `0' refers to quantities at present. Hence, the proper velocity of photons is not constant as in the standard FLRW metric, but it increases with the expansion as $a c$, where $c$ is the speed of light for $a = 1$. By contrast, the proper velocity of massive nonrelativistic particles is not affected by the universe expansion. This is in contradiction with behaviour of nonrelativistic massive particles in the standard FLRW metric, where the comoving velocity $v$ depends on $a$ as $a^{-2}$ and the proper velocity $V$ as $a^{-1}$.

\subsection{Conformal Friedmann equations }

Assuming the standard FLRW metric described by Eq. (A1), the Friedmann equation for the perfect isotropic fluid reads \citep{Peacock1999, Ryden2016}
\begin{equation}\label{eq14}
{\left({\frac{a'}{a}}\right)}^2 = \frac{8\pi G}{3} \rho - \frac{k c^2}{a^2} \,, 
\end{equation}
where $a'=da/dT$ is the derivative of the scale factor $a$ with respect to the proper time $T$, $G$ is the gravitational constant, $\rho$ is the mean mass density,  $k$ is the curvature index, and $k/a^2$ is the spatial curvature of the Universe. 

In order to express Eq. (14) in the conformal FLRW metric, we have to substitute the proper time $T$ by the conformal (comoving) time $t$ and time derivative $a' = da/dT$ by $\dot a = da/dt = a a'$. Hence, the conformal Friedmann equation reads
\begin{equation}\label{eq15}
{\left({\frac{\dot a}{a}}\right)}^2 = \frac{8\pi G}{3} \rho a^2 - k c^2 \,, 
\end{equation}
where $\dot a$ denotes the derivative with respect to the conformal time $t$. Considering the matter-dominated universe, we get
\begin{equation}\label{eq16}
\frac{8\pi G}{3} \rho =  H^2_0 \, \Omega_m a^{-3}  \,.
\end{equation}
Eq. (15) is rewritten as (see Vavry\v cuk \citep{Vavrycuk_Frontiers_2022})
\begin{equation}\label{eq17}
H^2\left(a\right) = H^2_0 \left({\Omega_m a^{-1} +  \Omega_k }\right) 
\end{equation}
with the condition
\begin{equation}\label{eq18}
\Omega_m + \Omega_k = 1 \,, 
\end{equation}
where $H(a) = \dot a/a$ is the Hubble parameter, $H_0$ is the Hubble constant,  $\Omega_m$ is the normalized matter density, and $\Omega_k$ is the normalized space curvature. Vavry\v cuk \citep{Vavrycuk_Frontiers_2022} proved that Eq. (17) describes the Type Ia supernova (SNe Ia) dimming well with no need to introduce dark energy, which is necessary in the standard $\Lambda$CDM model in order to fit the SNe Ia data.

Since this model is basically the Einstein-de Sitter (EdS) model but applied to the conformal FLRW metric, it will be called as the `conformal EdS model' in contrast to the standard EdS model based on the original FLRW metric.
 
Considering $a = 1/(1+z)$, the comoving time $t$ is expressed from Eq. (17) as a function of redshift as follows
\begin{equation}\label{eq19}
dt = \frac{1}{H_0 (1+z)} \left[\Omega_m (1+z) + \Omega_k\right]^{-1/2} dz \,, 
\end{equation}
and the proper time $T$ related to conformal $t$ as $dT = a(t) \, dt$ reads 
\begin{equation}\label{eq20}
dT = \frac{1}{H_0 (1+z)^2} \left[\Omega_m (1+z) + \Omega_k\right]^{-1/2} dz \,. 
\end{equation}
These relations are needed for relating observations of redshift to cosmic time.

\subsection{Gravitational orbits in the expanding universe}

Next, we study the influence of the expanding universe on the local gravity field produced by a point mass. The gravity field is assumed to produce a small perturbation of the metric tensor $g^{\mu\nu}$ describing the expanding space. So far, this problem has been studied under the assumption that the space expansion is defined by the standard FLRW metric (for a review, see Carrera and Giulini \citep{Carrera_Giulini_2010}) and the relevant equations are presented in Appendix A. Here, we derive equations for the gravitational orbits for the space expansion defined by the conformal FLRW metric. 

The homogeneous and isotropic expanding space characterized by the conformal FLRW metric will be disturbed by a spherically symmetric gravitational field produced by a point mass $M$ situated in the origin of coordinates. Since we limit to the weak gravitational effects of the point mass only, Eq. (5) is modified as follows
\begin{equation}\label{eq21}
\begin{split}
ds^2 = a^2(t) \left(-c^2(1+2\alpha)dt^2 + (1-2\alpha)\frac{dr^2}{1-kr^2}+r^2d\Omega^2\right)
\end{split}
\end{equation}
where 
\begin{equation}\label{eq22}
\alpha = -\frac{G M}{r c^2}\,, \, |\alpha| \ll 1
\end{equation}
is the Newtonian gravitational potential normalized to $c^2$, and $G$ is the gravitational constant. 

Let us assume a massive nonrelativistic particle ($v \ll c$) orbiting in the gravitational field in the plane defined by $\phi =0$. The metric tensor $g^{\mu \nu}$ is defined by Eq. (21). Calculating the Christoffel symbols $\Gamma^\mu_{\alpha \beta}$ in Eq. (9), we get
\begin{equation}\label{eq23}
\begin{gathered}
\ddot{r} - r {\dot{\theta}}^2 + \alpha c^2 + \frac{\dot{a}}{a} \dot{r} + \alpha\left( 2{\dot{\theta}}^2 r -3\frac{{\dot{r}}^2}{r} \right) \\
- \frac{1}{c^2} \frac{\dot{a}}{a} \dot{r} \left({\dot{r}}^2 + {\dot{\theta}}^2 r^2\right) = 0 \,, 
\end{gathered}
\end{equation}
\begin{equation}\label{eq24}
\begin{gathered}
r\ddot{\theta} + 2\dot{r}\dot{\theta} + \frac{\dot{a}}{a}r\dot{\theta} 
-2\alpha\dot{r}\dot{\theta} - 
\frac{1}{c^2} \frac{\dot{a}}{a} \dot{\theta} \left({\dot{r}}^2 + {\dot{\theta}}^2 r^2\right) = 0 \,, 
\end{gathered}
\end{equation}
where dots over quantities mean derivatives with respect to the conformal time $t$. Since $|\alpha| \ll 1$ and $c^2 \gg 1$, terms multiplied by $\alpha$ or by $1/c^2$ in Eqs (23) and (24) can be neglected and we get the following approximate equations
\begin{equation}\label{eq25}
\frac{1}{a} \frac{d}{dt} \left(a v^r \right) = -\frac{G M}{r^2} + \frac{\left(v^\theta\right)^2}{r} = f_g + f_c \,, 
\end{equation}
\begin{equation}\label{eq26}
\frac{1}{a} \frac{d}{dt} \left(a r v^\theta \right) = 0 \,, 
\end{equation}
where $v^r = \dot{r}$ is the radial comoving velocity, $v^{\theta} = r\dot{\theta}$ is the tangential comoving velocity, and $f_g$ and $f_c$ are the radial gravitational and centrifugal forces in the comoving coordinate system. If we assume that the orbit of the particle is stationary, the radial and centrifugal forces are balanced ($f_g = -f_c$) and the RHS of Eq. (25) equals zero. Consequently, we get
\begin{equation}\label{eq27}
a v^r = V^r = \mathrm{const} \,, 
\end{equation}
\begin{equation}\label{eq28}
a r v^\theta = r V^\theta = \mathrm{const} \,, 
\end{equation}
where $V^r$ and $V^\theta$ are the radial and tangential components of the proper velocity $V$. For a circular orbit in the comoving coordinates Eqs (27) and (28) are further simplified 
\begin{equation}\label{eq29}
V^r = 0,\, \, V^\theta = \mathrm{const} \,, 
\end{equation}
\begin{equation}\label{eq30}
r = \mathrm{const}\,, \,R = a R_0 \,, 
\end{equation}
where subscript `0' refers to the quantity at present.

\subsection{Physical consequences for the evolution of local systems}

Eq. (30) is surprising and in contradiction to the common opinion that the expansion of the Universe is without any appreciable effect on local gravitational systems \citep{Carrera_Giulini_2010}. This opinion is based on equations for gravitational orbits in an expanding space described by the standard FLRW metric (see Appendix A). The equations were derived by several authors \citep{Dicke_Peebles_1964, Pachner1964, Callan1965, Cooperstock1998, Faraoni_Jacques_2007, Sereno_Jetzer_2007, Carrera_Giulini_2010} and they differ from the standard Newtonian equations for orbits in the static universe only slightly. The only difference is that the equation for the radial acceleration of an orbiting body in the gravitational field imbedded in the space with the standard FLRW metric contains a term $\frac{\ddot{a}}{a}R$, which is related to the space expansion (see Appendix A, Eq. A7). It can be shown that this term might be appreciable in dynamics of large-scale structures as galaxy clusters, but the orbits within galaxies are affected negligibly. Consequently, all gravitationally bound systems with size of galaxies or smaller should be unaffected by the space expansion.

By contrast, considering the conformal FLRW metric for the expanding universe, the evolution of the local gravitational systems is essentially different from that obtained for the standard FLRW metric. In particular, the conformal FLRW metric predicts: 

\begin{enumerate}
\item 
An increase of the proper orbital radius $R_\mathrm{orb}$ with the expansion irrespective of the size of the local system and an increase of the proper orbital period $T_\mathrm{orb}$ with the expansion. Consequently, the size of galaxies (and all other local gravitational systems) must growth in the expanding Universe. The rate of growth is $(1+z)^{-1}$. The orbit is stationary in comoving coordinates but non-stationary in proper coordinates.
\item 
A constant proper rotation velocity $V$ of particles along the orbits irrespective of the orbital radius during the space expansion. Hence, the stationary circular orbits in the comoving coordinate system become spirals in the proper coordinate system. Stars and gas within spirals have flat rotation curves. Since flat rotation curves are a consequence of the expansion of the Universe, no dark matter is needed to explain dynamics of galaxies. 
\end{enumerate}

The properties of spiral galaxies and their evolution in cosmic time are exemplified by numerical modelling in the next section. 

\section{Numerical modelling of the galaxy dynamics}

\begin{figure*}
\includegraphics[angle=0,width=16 cm, clip=true, trim=60 190 80 140]{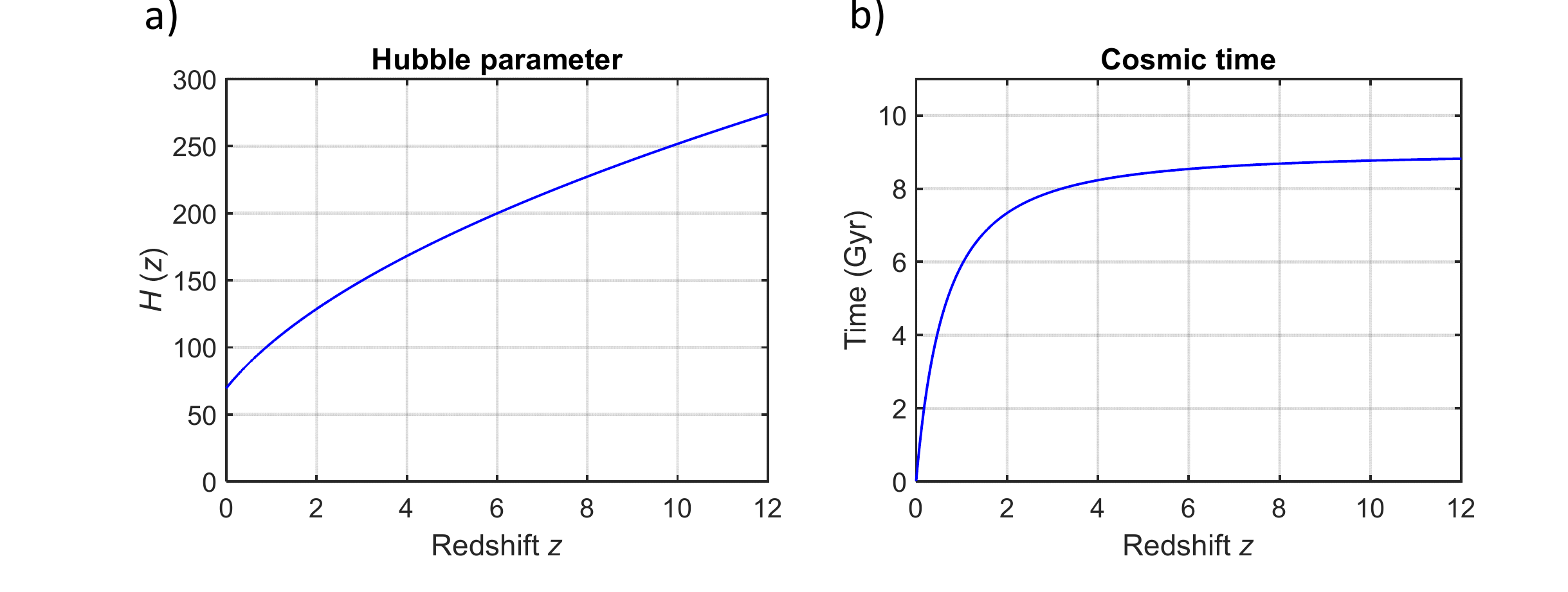}
\caption{
The Hubble parameter (a) and the proper cosmic time (b) as a function of redshift. The cosmological model is described by Eq. (17) for the conformal FLRW metric with the with $\Omega_m = 1.2$, and  $\Omega_k = -0.2$ (see Vavry\v cuk \citep{Vavrycuk_Frontiers_2022}). The Hubble constant is $H_0 = 69.8 \, \mathrm{km \, s^{-1} Mpc^{-1}}$, obtained by Freedman et al. \citep{Freedman2019} from observations of the SNe Ia data with a red giant calibration.
}
\label{fig:1}
\end{figure*}

\begin{figure*}
\includegraphics[angle=0,width=16 cm, clip=true, trim=60 190 80 140]{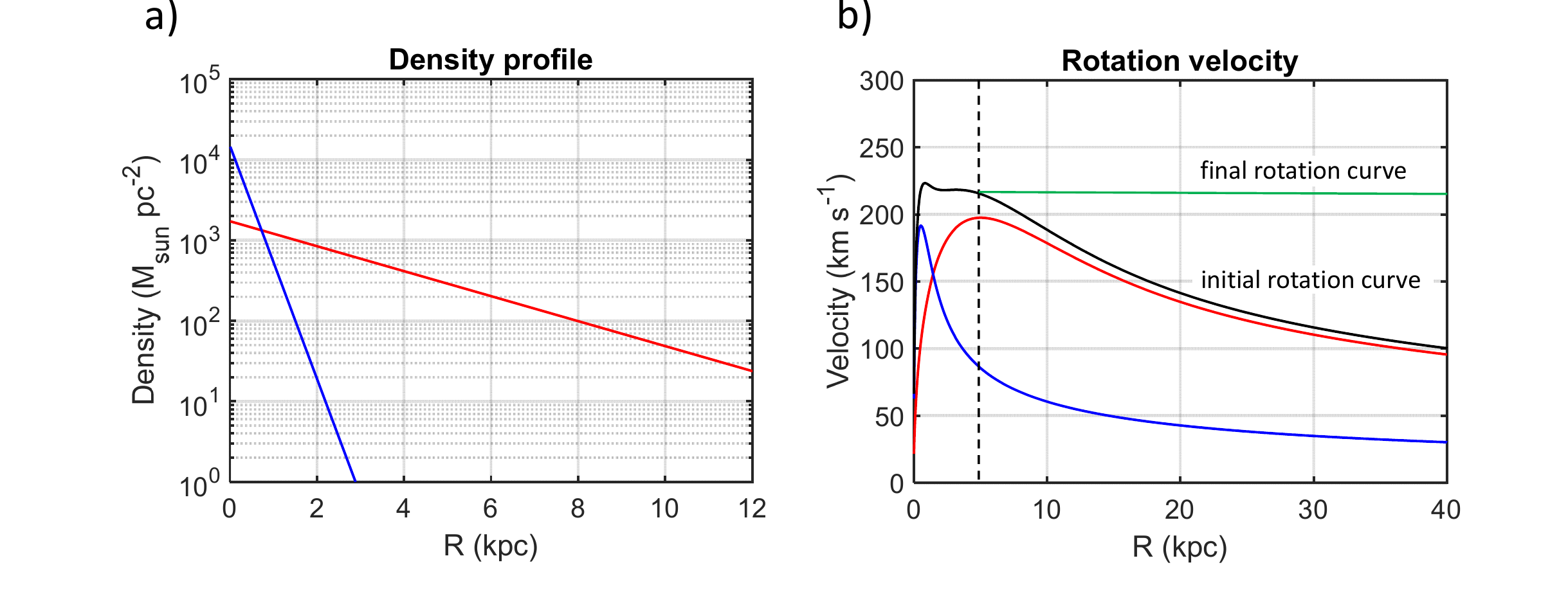}
\caption{
The density profile (a) and the initial and final rotation velocity curves (b) for a simulated galaxy. The surface densities of the bulge (blue line in a) and disk (red line in a) are calculated by Eqs (31) and (32) with $\Sigma_b (R_{b0}) = 3.6 \, \mathrm{M_\odot \, pc^{-2}}$, $R_{b0} = 2.5 \, \mathrm{kpc}$, $R_b = 0.3 \, \mathrm{kpc}$ for the bulge, and with $\Sigma_d (R_{d0}) = 99.2 \, \mathrm{M_\odot \, pc^{-2}}$, $R_{d0} = 8.0 \, \mathrm{kpc}$, $R_d = 2.8 \, \mathrm{kpc}$ for the disk. The total masses of the bulge and disk are $M_b = 8.5 \times 10^9 \, \mathrm{M_\odot}$ and $M_d = 8.5 \times 10^{10} \, \mathrm{M_\odot}$, respectively. The total initial rotation velocity (black line in b) is shown together with the contribution of the bulge (blue line in b) and disk (red line in b). The dashed line in (b) defines the boundary $R_c = 5 \, \mathrm{kpc}$ between the bulge-bar regime and the trailing-spiral regime. The green line in (b) shows the final rotation curve caused by the space expansion (the stellar mass and gas with $R > R_c$ are continuously moving out of the galaxy centre, but they keep their rotation velocity).
}
\label{fig:2}
\end{figure*}

\subsection{Parameters for modelling}

In this section, we will examine the evolution of local gravitationally bounded systems in the expanding universe by numerical modelling. As an example, we will focus on the evolution of a spiral galaxy with parameters close to the Milky Way. For modelling, we need to specify the expanding history of the Universe described by the Hubble parameter. Since we do not use the standard FLRW metric, we cannot adopt the $\Lambda$CDM model. Instead, the conformal FLRW metric described by Eq. (17) must be used for deriving the Hubble parameter. According to Vavry\v cuk \citep{Vavrycuk_Frontiers_2022}, we use in Eq. (17) parameters $\Omega_m = 1.2$, $\Omega_\Lambda = 0$ (no dark energy), and $\Omega_k = -0.2$ (closed universe). The Hubble constant is $H_0 = 69.8 \,\mathrm{km\, s^{-1}\, Mpc^{-1}}$, obtained by Freedman et al. \citep{Freedman2019} from observations of the SNe Ia data with a red giant calibration. Obviously, Eqs (17) and (19) predict a quite different evolution of the Hubble parameter and a time-redshift relation than the $\Lambda$CDM model. For example, redshift $z$ of 4.12 corresponds to the cosmic time of 15 Gyr (see Fig. 1).

The galaxy is assumed to be formed by a bulge and disk with the following exponential density profiles \citep{McGaugh2016_ApJ} 
\begin{equation}\label{eq31}
\Sigma_b(R) = \Sigma_b(R_{b0})e^{-(r-R_{b0})/R_b} \,, 
\end{equation}
\begin{equation}\label{eq32}
\Sigma_d(R) = \Sigma_d(R_{d0})e^{-(r-R_{d0})/R_d} \,, 
\end{equation}
where $\Sigma_b(R)$ and $\Sigma_d(R)$ are the surface densities of the bulge and disk, respectively. The required parameters are for the bulge $\Sigma_b(R_{b0})= 3.6 \, \mathrm{M_\odot \, pc^{-2}}$, $R_{b0}=2.5 \, \mathrm{kpc}$, $R_b = 0.3 \, \mathrm{kpc}$, and for the disk $\Sigma_d(R_{d0}) = 99.2 \, \mathrm{M_\odot \, pc^{-2}}$, $R_{d0} = 8.0 \, \mathrm{kpc}$, $R_d = 2.8 \, \mathrm{kpc}$. The total masses of the bulge and disk are $M_b = 8.5 \times 10^9 \, \mathrm{M_\odot}$ and $M_d = 8.5 \times 10^{10} \, \mathrm{M_\odot}$, respectively. The bulge is usually modelled as a prolate, triaxial bar \citep{Binney1997, Bissantz_Gerhard_2002}, but we do not focus on modelling a 3D geometry of the bulge and its evolution, so the simplified approximation described by Eq. (31) is satisfactory.

\begin{figure*}
\includegraphics[angle=0,width=16 cm, clip=true, trim=120 190 120 160]{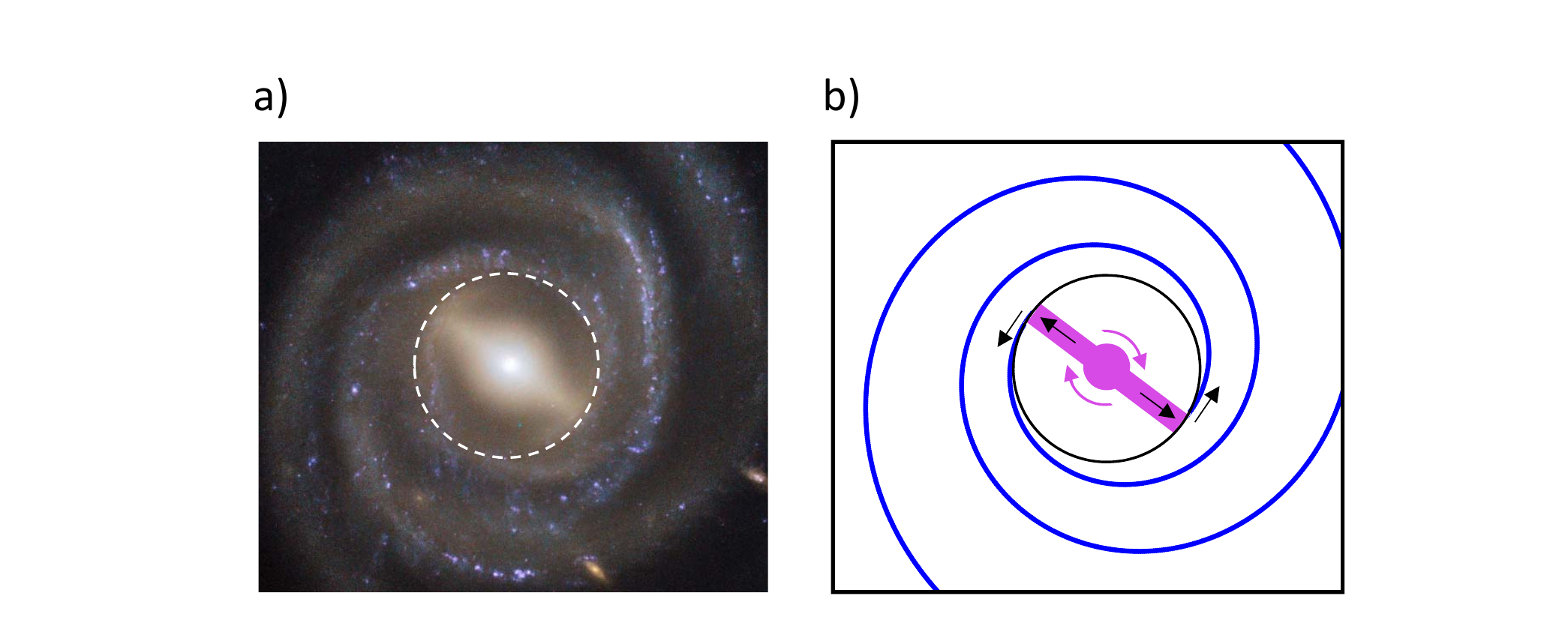}
\caption{
The spiral galaxy UGC 6093 observed by ESA/Hubble (a) and the scheme of the spiral galaxy (b). The bulge and bar of the galaxy in plot (b) is marked by the purple colour. The purple arrows in (b) show the rotation direction of the bar. The black arrows in (b) show the direction of the outflow from the bulge-bar domain into the spiral domain. The white dashed circle in (a) and the solid black circle in (b) mark the boundary between the bulge-bar and spiral domains.
}
\label{fig:3}
\end{figure*}
\begin{figure*}
\includegraphics[angle=0,width=16 cm, clip=true, trim=60 190 80 140]{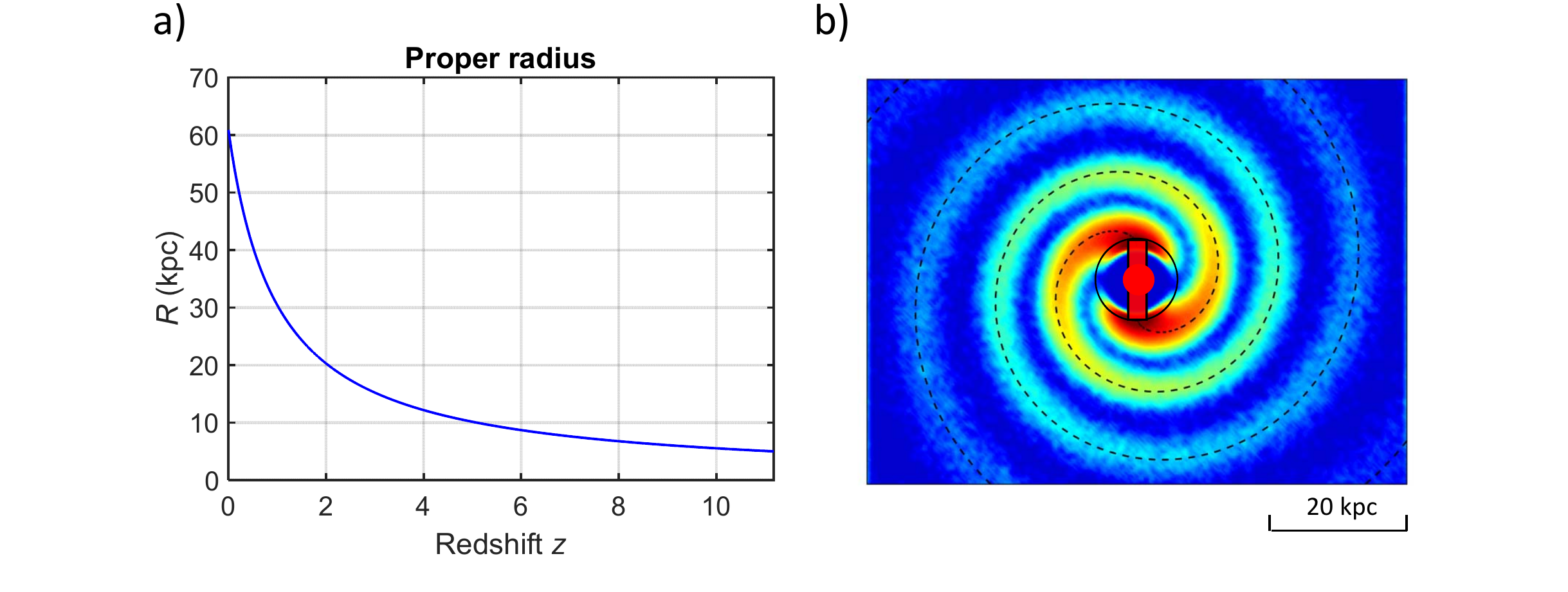}
\caption{
The proper radius of a galaxy as a function of redshift (a) and the spiral arms formed during the galaxy evolution (b). The age of the galaxy is 8.8 Gyr and the maximum redshift is $z = 11.2$. The solid black circle in (b) marks the boundary between the bulge-bar and spiral domains with radius $R_c = 5 \, \mathrm{kpc}$. The black dashed line in (b) denotes the central orbit with unperturbed parameters $R_c$ and $\theta$. The other orbits forming the spirals are characterized by perturbed parameters $R_c$ and $\theta$, see the text. The red bulge and bar inside the black circle in (b) is a scheme illustrating the orientation of the bar.
}
\label{fig:4}
\end{figure*}

\subsection{Bulge-bar region and spirals }

The density profiles and the Newtonian rotation curves predicted for the used parameters are shown in Fig. 2. The total rotation curve (Fig. 2b, black line) is separated into two domains. The separation (critical) distance $R_c$ is about 5 kpc (Fig. 2b, dashed vertical line) and corresponds to the maximum rotation velocity associated with the disk density profile (Fig. 2b, red line). For shorter distances (bulge-bar domain), the behaviour of the rotation curve is complex being affected by both the bulge and disk. For larger distances (spiral domain), the initial rotation curve is monotonously decreasing affected mostly by disk only. Obviously, the critical distance $R_c$ may vary for different galaxies being dependent on their mass and density profiles. As discussed below, the rotation curve evolves in time and the final rotation curve becomes flat in the spiral domain (Fig. 2b, green line).

Fig. 3 shows the bulge-bar and spiral regions for the spiral galaxy UGC 6093 together with a scheme suggesting a possible origin of spirals and their temporal evolution. The bulge-bar region is characterized by a high concentration of stars and gas. Consequently, gravitational forces maintain the shape of the galaxy inside this region, irrespective of its rotation. Since the bar is continuously increasing due to the space expansion, its ends cross the region boundary and outflow from the bulge-bar region into the spiral region. In this region, the radial gravitational forces become dominant and stars and gas move as bodies in a spherically symmetric gravitational field. Their motion will be described by Eqs (29) and (30): the rotation velocity will be constant with time but the orbital radius will increase. Consequently, the rotating bar will outstrip the material in the spiral region, which will form a pattern of trailing spirals. The material in all spirals will display flat rotation curves and the size of galaxies will increase with time. 

\subsection{Scenarios of the galaxy evolution}

\begin{figure*}
\includegraphics[angle=0,width=16 cm, clip=true, trim=60 135 80 60]{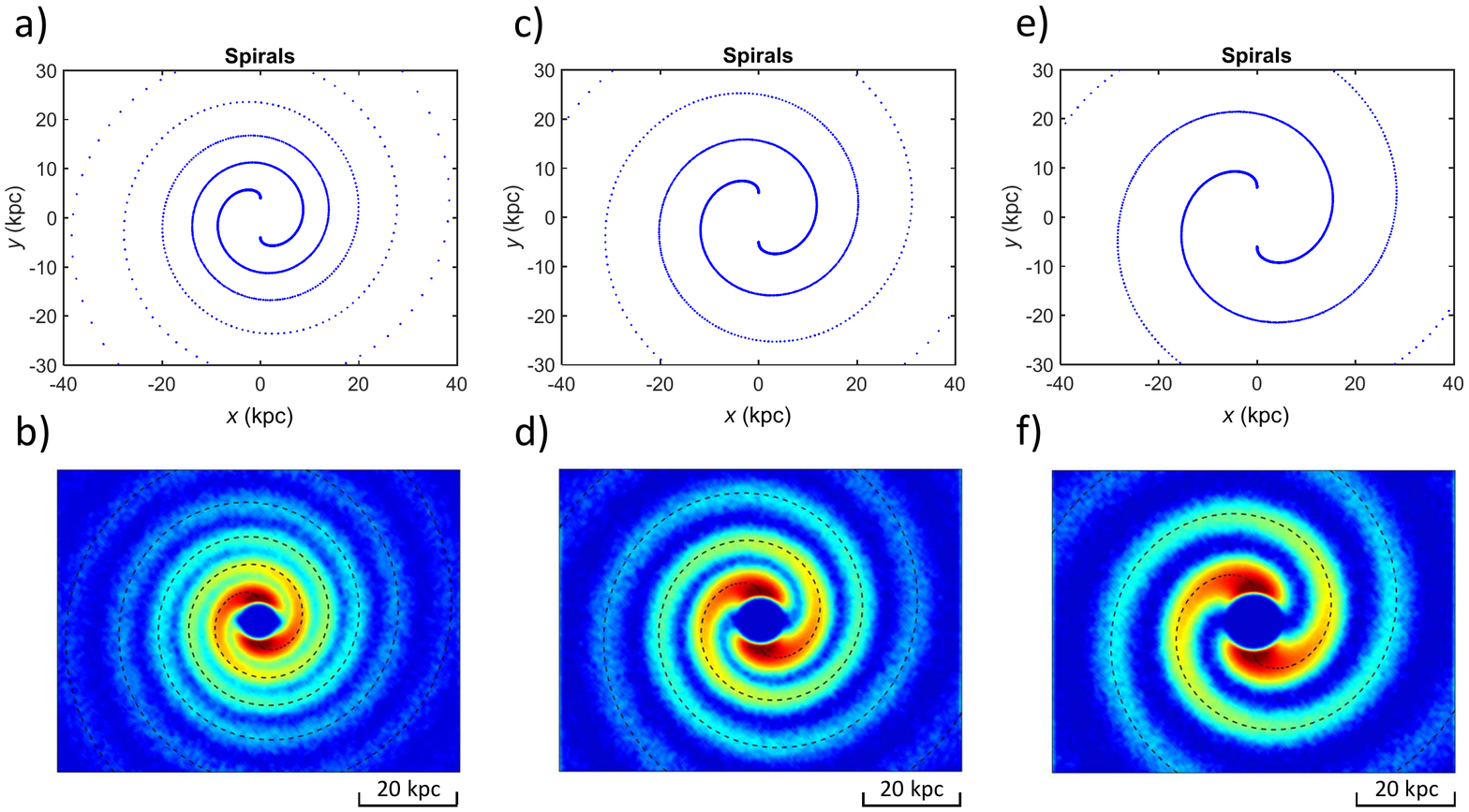}
\caption{
Geometry of spirals: dependence on the initial size of the bulge-bar region. The age of the galaxy is 8.8 Gyr and the maximum redshift is $z = 11.2$. The upper/lower plots show orbits with unperturbed/perturbed parameters $R_c$ and $\theta$. The critical radius $R_c$ and the rotation velocity $V^\theta$ are 4 kpc and 222.7 km/s in (a,b), 5 kpc and 220.1 km/s in (c,d), and 6 kpc and 215.9 km/s in (e,f). The masses of the bulge and disk are $M_b = 8.5 \times 10^9 \, \mathrm{M_\odot}$ and $M_d = 8.5 \times 10^{10} \, \mathrm{M_\odot}$, respectively. 
}
\label{fig:5}
\end{figure*}

\begin{figure*}
\includegraphics[angle=0,width=16 cm, clip=true, trim=60 135 80 60]{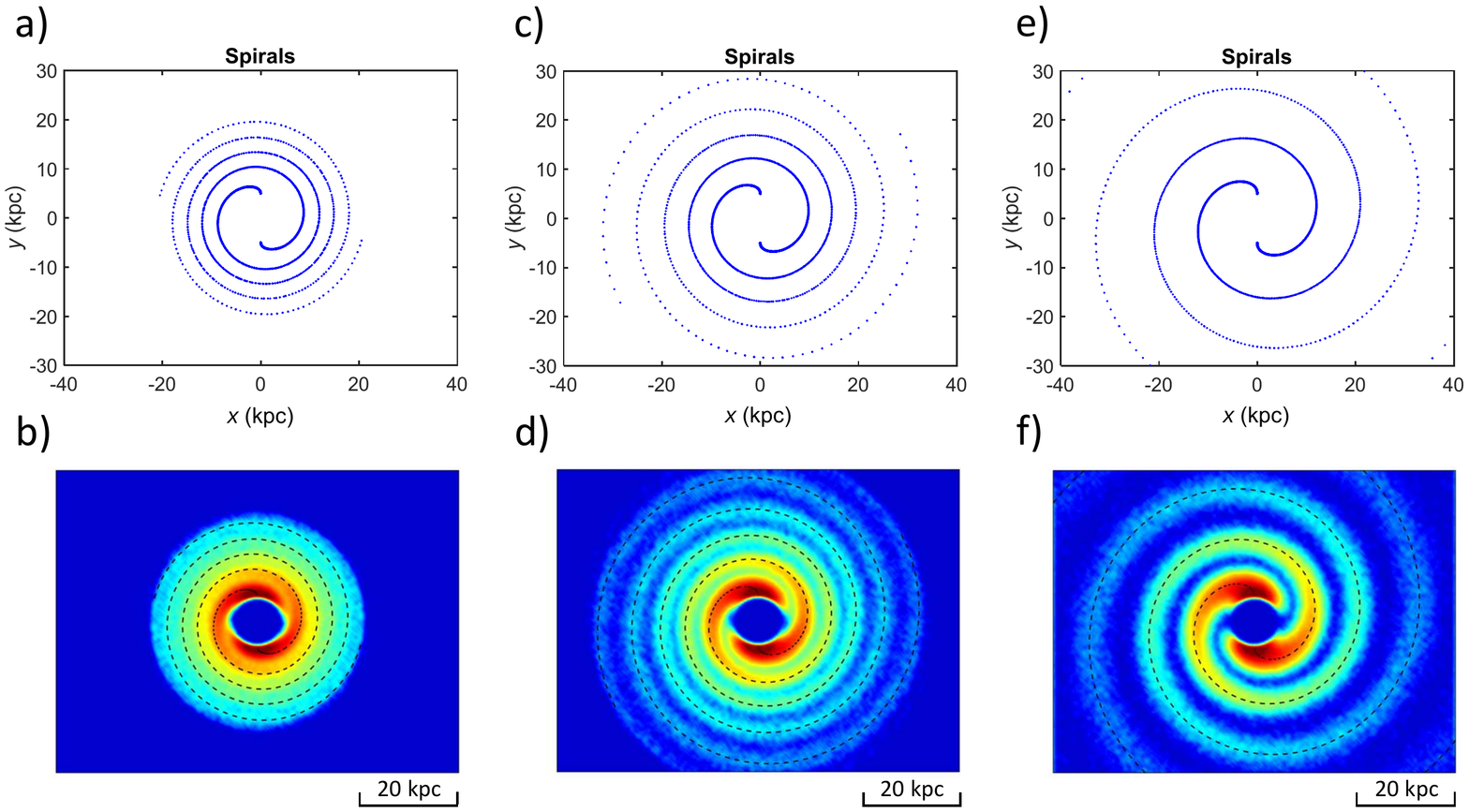}
\caption{
Geometry of spirals: dependence on the galaxy age. The critical radius $R_c$ and the rotation velocity $V^\theta$ are 5 kpc and 220.1 km/s. The upper/lower plots show orbits with unperturbed/perturbed parameters $R_c$ and $\theta$. The galaxy age and the maximum redshift are 8.0 Gyr and $z = 3.20$ in (a,b), 8.5 Gyr and $z = 5.65$ in (c,d), and 8.8 Gyr and $z = 11.18$ in (e,f). The masses of the bulge and disk are $M_b = 8.5 \times 10^9 \, \mathrm{M_\odot}$ and $M_d = 8.5 \times 10^{10} \, \mathrm{M_\odot}$, respectively. 
}
\label{fig:6}
\end{figure*}

\begin{figure*}
\includegraphics[angle=0,width=16 cm, clip=true, trim=60 135 80 60]{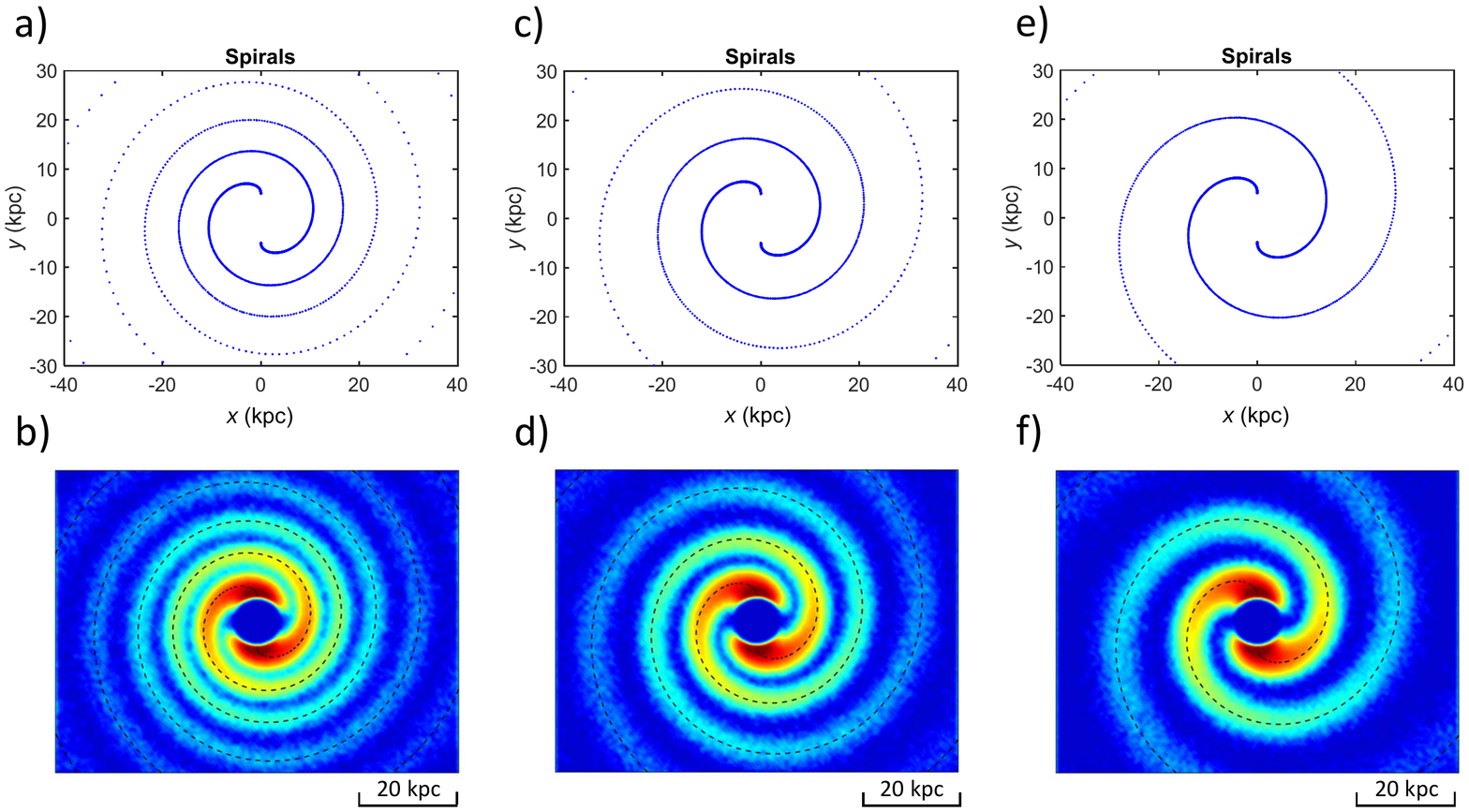}
\caption{
Geometry of spirals: dependence on the galaxy mass. The critical radius $R_c$, the galaxy age and the maximum redshift are 5 kpc, 8.8 Gyr and $z = 11.18$. The upper/lower plots show orbits with unperturbed/perturbed parameters $R_c$ and $\theta$. The masses of the bulge and disk are $M_b = 17 \times 10^9 \, \mathrm{M_\odot}$ and $M_d = 17 \times 10^{10} \, \mathrm{M_\odot}$ in (a,b), $M_b = 8.5 \times 10^9 \, \mathrm{M_\odot}$ and $M_d = 8.5 \times 10^{10} \, \mathrm{M_\odot}$ in (c,d), and $M_b = 4.25 \times 10^9 \, \mathrm{M_\odot}$ and $M_d = 4.25 \times 10^{10} \, \mathrm{M_\odot}$ in (e,f). The rotation velocity $V^\theta$ is 311.3 km/s in (a,b), 220.1 km/s in (c,d), and 155.6 km/s in (e,f). 
}
\label{fig:7}
\end{figure*}

The form of spirals depends on several factors: (1) the radius of the bulge-bar area, (2) the age of the galaxy, (3) the mass of the galaxy, density profile and its rotation velocity, and (4) the expansion history of the universe. Therefore, we assume several alternative scenarios in the numerical modelling, where we vary some of these parameters. The radius of the bulge-bar area $R_c$ will be considered 4, 5 and 6 kpc. The age of a galaxy will be 8.0, 8.5 and 8.8 Gyr. This age will correspond to the redshift $z$ of 3.2, 5.6 and 11.2. The masses of the bulge and disk of a galaxy are $M_b = 8.5 \times 10^9 \, \mathrm{M_\odot}$ and $M_d = 8.5 \times 10^{10} \, \mathrm{M_\odot}$, but we will also model a galaxy with a half of this mass and with a twice higher mass.

In order to simulate a more realistic galaxy evolution, we assume random conditions for the bulge-bar outflow. The critical distance $R_c$ is not defined by a single value, but it varies according to the Gaussian distribution with the standard deviation of 0.1 kpc. Similarly, the bulge-bar outflow does not cross the domain boundary at two single points defined at the two ends of the bar by angles $\theta = 0^{\circ}$ and $180^{\circ}$. Instead, angle $\theta$ obeys the Gaussian distribution centred at $0^{\circ}$ and $180^{\circ}$ with the standard deviation of $20^{\circ}$.

\subsection{Results}

In modelling, we calculate orbits of stars and gas flowed out from the bulge-bar domain into the spiral domain and evolution of orbits in time. Fig. 4 shows the evolution of a galaxy with age of 8.8 Gyr. The galaxy started to evolve at redshift $z$ of 11.2. The galaxy was formed just by the bulge and bar with no spirals at the beginning of the simulation. At this time, the radius of the galaxy was 5 kpc being the same as the radius of the bulge-bar area. The material, which outflowed from the bulge-bar domain, formed typical trailed spirals during the galaxy evolution (Fig. 4b). Hence, the radius of the galaxy increased from 5 kpc to 61 kpc during its life (Fig. 4a). The rotation velocity of the material in the spirals is uniform and attains a value of 220 km/s.

Fig. 5 demonstrates a dependence of the spiral pattern on the size of the bulge-bar area. The age of the galaxy is again 8.8 Gyr and the evolution started at redshift $z$ of 11.2. For a smaller radius of the bulge-bare area (Fig. 5a, $R_c = 4 \, \mathrm{kpc}$), the rotation velocity is slightly higher being 223 km/s. As a result, the orbital period is smaller and the recession of the spirals from the central part of the galaxy is not so pronounced. By contrast, for a larger radius of the bulge-bare area (Fig. 5c, $R_c = 6 \, \mathrm{kpc}$), the rotation velocity is slightly lower being 216 km/s. The orbital period is higher and the recession of the spirals from the central part of the galaxy is more distinct.

Fig. 6 shows how is the spiral pattern affected by the age of the galaxy. The figure shows three galaxies with the age of 8.0, 8.5 and 8.8 Gyr. The corresponding redshifts are 3.2, 5.6 and 11.2. The radius of the bulge-bar area is $R_c = 5 \, \mathrm{kpc}$ and the rotation velocity is 220 km/s for all three galaxies. As expected, the younger the galaxy, the less evolved the spiral pattern. Consequently, the final radius of the galaxy increased from 5 kpc to 21, 33 and 61 kpc, respectively.

Finally, Fig. 7 presents how is the spiral pattern affected by the mass of a galaxy. Fig. 7b shows a galaxy with masses of the bulge and disk of a galaxy are $M_b = 8.5 \times 10^9 \, \mathrm{M_\odot}$ and $M_d = 8.5 \times 10^{10} \, \mathrm{M_\odot}$ (rotation velocity at $R_c = 5 \, \mathrm{kpc}$ is 220 km/s). These values were used in all previous simulations (Figs 4-6). Fig 7a shows a galaxy, which has a twice higher mass. The corresponding rotation velocity at $R_c = 5 \, \mathrm{kpc}$ is 311 km/s. By contrast, Fig 7c shows a galaxy which has a twice lower mass. The corresponding rotation velocity at $R_c = 5 \, \mathrm{kpc}$ is 156 km/s only. The age of the three galaxies is 8.8 Gyr and their evolution started at redshift $z = 11.2$. Since the galaxies have the same age, their size increases in the same way: from 5 kpc to 61 kpc. Still the spiral pattern is remarkably different for all three galaxies. The massive galaxy rotates fast and the spirals are receding slowly from the galaxy centre (Fig 7a). By contrast, the galaxy with the smallest mass rotates slowly and the recession of spirals from the galaxy centre is high (Fig 7c).

\section{Supporting observational evidence}
The presented results are supported by many observations that are difficult to explain under the standard cosmological model. This applies to galaxy dynamics, morphology of spiral galaxies as well as dynamics of the solar system. In the next, we review several puzzles in modern cosmology resolved by the proposed theory.

\subsection{Galaxy expansion}
\begin{figure*}
\includegraphics[angle=0,width=16 cm, clip=true, trim=60 155 80 120]{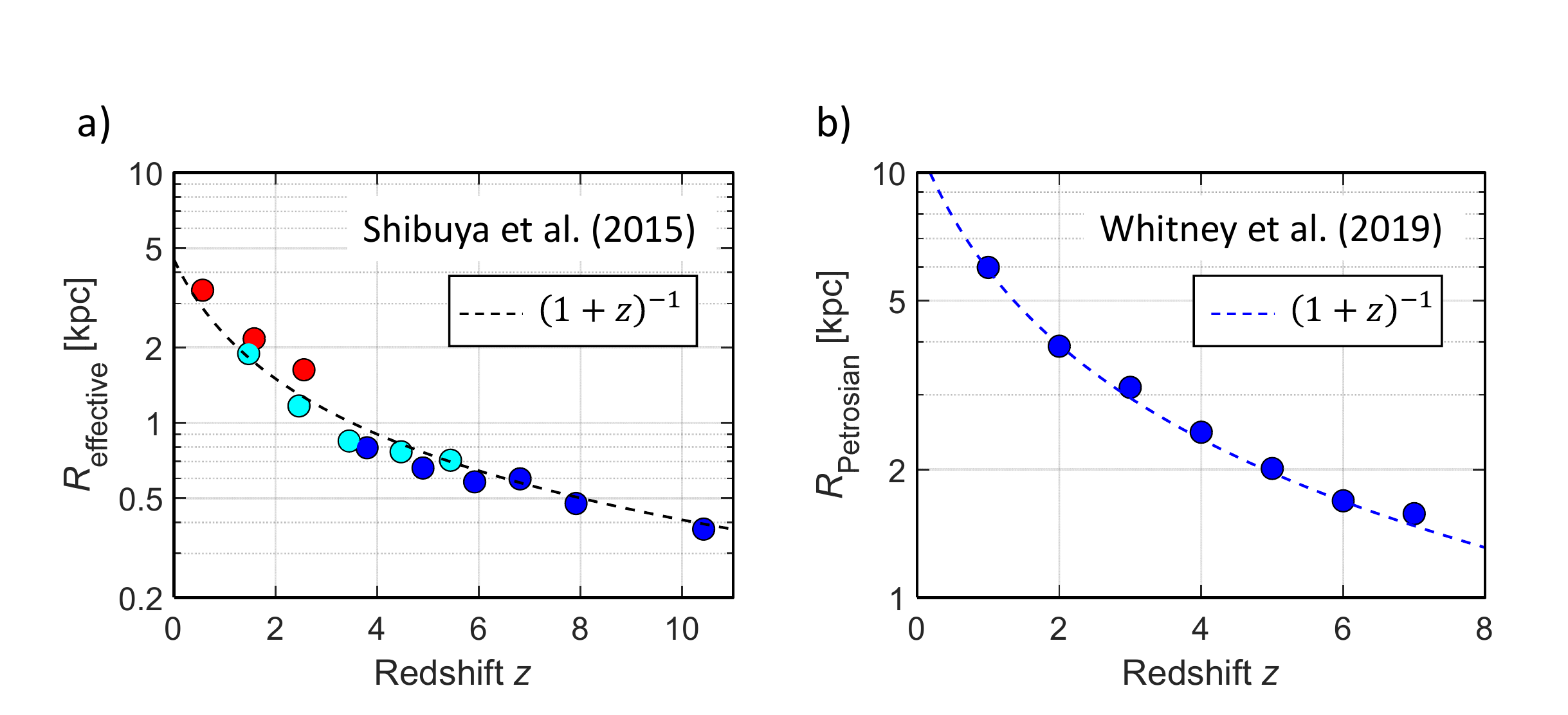}
\caption{
Galaxy size evolution with redshift. (a) Median of the effective galaxy radius $R$ as a function of redshift for galaxies in the bin of $L_{UV} = (0.3-1) \, L_{z=3}^*$. The red and cyan filled circles indicate radius $R$ for star-forming galaxies measured in the optical ($4500 - 8000 \,$\AA) and UV ($1500 - 3000 \,$\AA) wavelength ranges, respectively. The blue filled circles indicate radius $R$ for the Lyman break galaxies measured in the UV  wavelength range. For details, see Shibuya et al. \citep{Shibuya2015}, their fig. 8. (b) Median Petrosian radius of galaxies as a function of redshift for the mass-limited sample in the range of $10^9 \, \mathrm{M_\odot} \le M_* \le 10^{10.5} \, \mathrm{M_\odot}$. The ratio of the surface brightness at radius $R$ to the mean surface brightness of a galaxy is $\eta = 0.2$. For details, see Whitney et al. \citep{Whitney2019}, their fig. 8. The dashed lines in (a,b) show the size evolution predicted by the presented theory.
}
\label{fig:8}
\end{figure*}

As shown in the previous sections, the size of galaxies should increase with redshift as $(1+z)^{-1}$ with no change of the galaxy mass. Based on observations, it is commonly accepted that the size of galaxies evolves rapidly during the cosmic time \citep{vanDokkum2008, vanDokkum2010, Williams2010}, see Fig. 8. Using observations from the Hubble Space Telescope (HST), galaxy sizes defined by the effective radius, $R_e$, have been extensively measured with the Advanced Camera for Surveys (ACS) and the Wide Field Camera 3/IR channel on board HST for massive galaxies at $z < 3$ (e.g., van der Wel et al. \citep{van_der_Wel2014}) and $z \ge 3-4$ Lyman break galaxies (LBGs) selected in the dropout technique \citep{Trujillo2006, Dahlen2007, McLure2013a}. The average size is reported to evolve according to $R_e \sim (1 + z)^{-B}$ , with $B$ ranging most frequently between 0.8 and 1.2 \citep{Oesch2010, Bouwens2004, Holwerda2015}. For example, Shibuya et al. \citep{Shibuya2015} studied the redshift evolution of the galaxy effective radius $R_e$ obtained from the HST samples of $\sim$190,000 galaxies at $z = 0-10$, consisted of 176,152 photo-$z$ galaxies at $z = 0-6$ from the 3D-HST+CANDELS catalogue and 10,454 Lyman break galaxies (LBGs) at $z = 4-10$ identified in the CANDELS, HUDF 09/12, and HFF parallel fields. They found that $R_e$ values at a given luminosity decrease toward high $z$, as $R_e \sim (1 + z)^{-B}$, with $B = 1.10 \pm 0.06$ for median, see Fig. 8a.

Since it is believed that the size of galaxies cannot be affected by the expansion of the Universe, the observed expansion of galaxies is explained by other mechanisms. The most popular theory suggests the growth of galaxies being produced by galaxy mergers \citep{Naab2009, McLure2013a, Kormendy_Ho_2013, Conselice2014}. An important role in  merging of galaxies play dark matter haloes \citep{Kauffmann1993, Mo1998}. Two types of mergers are distinguished: a major merger where the stellar masses of the galaxies are comparable, and a minor merger where the stellar mass of one galaxy is much lower. 

However, the idea of the galaxy expansion due to galaxy mergers is controversial for several reasons \citep{Lerner2018}. First, observations indicate that the major and minor merger rates are much lower to explain the galaxy expansion \citep{Taylor2010, Man2012, Man2016}. For example, Mundy et al. \citep{Mundy2017} report approximately 0.5 major mergers at $z < 3.5$ representing an increase in stellar mass of 20-30\% only when considering constant stellar mass samples. As regards minor mergers, Newman et al. \citep{Newman2012} studied 935 galaxies selected with $0.4 < z < 2.5$ and concluded that minor merging cannot account for a rapid growth of the size seen at higher redshifts. Man et al. \citep{Man2016} studied massive galaxies using the UltraVISTA/COSMOS catalogue, complemented with the deeper, higher resolution 3DHST+CANDELS catalogue and estimated $\sim$1 major merger and $\sim$0.7 minor merger on average for a massive ($M_* \ge 10^{10.8} \, \mathrm{M_\odot}$) galaxy during $z = 0.1-2.5$. The observed number of major and minor mergers can increase the size of a massive quiescent galaxy by a factor of two at most. Hence,  additional mechanisms are needed to fully explain the galaxy evolution. Second, mergers cannot explain the growth of spiral galaxies, because mergers destroy disks as shown by Bournaud et al. \citep{Bournaud2007}. Third, the idea of mergers implies an increase of stellar mass in galaxies over cosmic time. However, observations show no or slight mass evolution in time \citep{Bundy2017, Kawinwanichakij2020}.

\subsection{Galaxy rotation curves}

Another basic characteristics predicted by the presented theory are flat rotation curves of spiral galaxies. In Newton theory, a velocity of stars in a rotating galaxy is controlled by gravitational and centrifugal forces only. Assuming the Newton's gravitation law, a balance between the forces implies a decay of the orbital speed $V(R)$ of a star with its distance $R$ from the galaxy centre 
\begin{equation}\label{eq33}
V^2(R) = \frac{G M (R)}{R} \,, 
\end{equation}
where $M(R)$ is the mass of the galaxy as a function of $R$. Hence, the rotation curve $V(R)$ decays as $R^{-1/2}$ provided the most of mass is concentrated in the galaxy centre. The same applies to the improved Newtonian equations in an expanding Universe described by the standard FLRW metric. Obviously, the flat rotation curves predicted by the improved Newtonian equations in the Universe with the conformal FLRW metric point to fundamental differences between both the metrics. Importantly, the flat rotation curves of spiral galaxies are observationally confirmed.

\subsubsection{Observations of flat rotation curves}

\begin{figure}
\includegraphics[angle=0,width=8 cm, clip=true, trim=80 60 80 40]{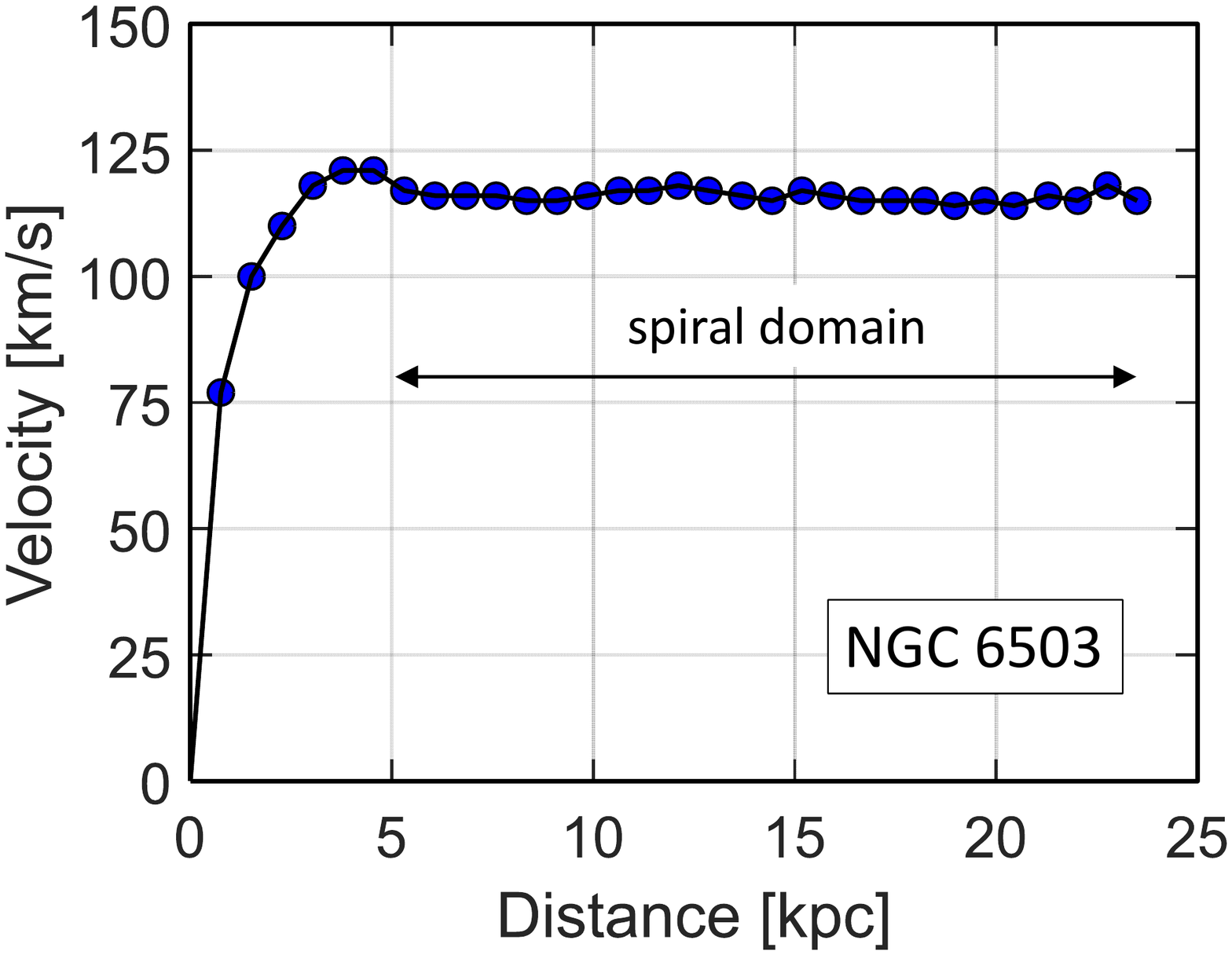}
\caption{
Rotation curve for the NGC 6503 galaxy. The spiral domain covering distances greater than 5 kpc is indicated. In this domain, receding of spirals from the galaxy centre causes the flatness of the rotation curve. For data, see Begeman \citep{Begeman1987} and Lelli et al. \citep{Lelli2016a}.
}
\label{fig:9}
\end{figure}

Rubin and Ford \citep{Rubin_Ford1970} discovered that the rotation curve of the Andromeda galaxy has a sharp maximum of $V = 225 \, \mathrm{km/s}$ at $R = 400 \, \mathrm{pc}$, a deep minimum at $R =2 \, \mathrm{kpc}$ and at $R > 3 \, \mathrm{kpc}$ it is nearly flat with the maximum velocity of $270 \pm 10 \, \mathrm{kpc}$. Such behaviour was later confirmed also for other spiral galaxies \citep{Rubin1980, Rubin1985, Albada1985, Begeman1989, Sanders1996, Swaters2000, Sofue_Rubin2001, deBlok2002, deBlok2008, McGaugh2019, Tiley2019}, for an example, see the rotation curve of the NGC 6503 galaxy in Fig. 9. The measurements of rotation curves are mostly based on \citep{Sofue2017}: observations of emission lines at optical wavelengths such as H$\alpha$ and [N$_\mathrm{II}$] lines, particularly, in H$_\mathrm{II}$ regions in galactic disks; at infrared wavelengths revealing kinematics of dusty disks and nuclear regions of spiral galaxies with significant dust extinction; and at 21-cm H$_\mathrm{I}$ line powerful to study kinematics of entire spiral galaxy.

The rotation curves of spiral galaxies display a significant similarity irrespective of their morphology \citep{Persic1996, Sofue2016, Sofue2017}. The differences are mainly connected to the mass and size of the galaxies. More massive and larger galaxies (Sa and Sb) have high rotation velocity close to the nucleus, while smaller galaxies (Sc) show slower rotation in the centre. The earlier-type (Sa and Sb) galaxies display a flat or slowly declining velocity at the outermost part of the rotation curve, while the rotation velocity of the later-type (Sc) galaxies monotonically increases. Similarly, dwarf and LSB galaxies display monotonically increasing rotation velocity until their galaxy edges \citep{Swaters2000, Swaters2003, deBlok2001}. In addition, Tully and Fisher \citep{Tully_Fisher_1977} revealed an empirical statistical relation between the galaxy luminosity and the maximum rotation velocity at a few galactic disk radii. The Tully-Fisher relation is commonly used for estimating the luminosity of distant galaxies \citep{Mathewson1992, Jacoby1992, Phillips1993} and for measuring the Hubble constant \citep{Mould2000, Freedman2001}.

\subsubsection{Dark matter}

To explain the discrepancy between predicted and observed rotation curves, several theories have been proposed. The most straightforward way is to assume the presence of dark matter (DM) with distribution calculated as (Schneider \citep{Schneider2015}, his eq. 3.17)
\begin{equation}\label{eq34}
M_{\mathrm{dark}} (R) = \frac{R}{G}\left(V^2_{\mathrm{obs}} (R) - V^2 (R) \right) \,, 
\end{equation}
where $V_{\mathrm{obs}} (R)$ is the observed rotation velocity and $V(R)$ is calculated according to Eq. (33). The idea of dark matter origins from Zwicky \citep{Zwicky1933,Zwicky1937} who postulated `missing mass' to account for the orbital velocities of galaxies in clusters. Originally, the DM was assumed to be baryonic formed by gas, dust and microscopic and macroscopic solid bodies including black-holes. Later on, the baryonic origin of DM was questioned and rejected. Since the analysis of galaxy rotation curves revealed that the mass of DM \citep{Albada1985, Dubinski_Carlberg1991, Navarro1996, Navarro1997, Persic1996} is much higher than estimates of dust and gas in galaxies \citep{Calzetti2000, Dunne2000, Draine2007, Cunha2008, Sandstrom2013}, DM was considered to be mostly of non-baryonic nature \citep{White_Rees1978, Davis1985, White1987, Maddox1990a, Moore1999, Bergstrom2000, Bertone_Hooper2018}.  To reconcile the theoretical and observed rotation curves, the DM is significant at large distances and forms a DM halo which total mass exceeding the stellar galaxy mass by about one order or more \citep{Albada1985, Dubinski_Carlberg1991, Navarro1996, Persic1996}.

\subsubsection{MOND theory}

The non-baryonic DM concept is not, however, unanimously accepted. The non-baryonic DM is questioned, in particular, for its exotic and mysterious nature and for difficulties to be detected by other methods independent of gravity. Also, significant discrepancies with predictions of the $\Lambda$CDM model on small scale are reported by many authors \citep{Kroupa2015, Del-Popolo_Le-Delliou2017}. Therefore, several alternative theories have been proposed to explain the flat rotation curves \citep{Mannheim2006}. The most famous alternative theory was presented by Milgrom \citep{Milgrom1983a,Milgrom1983b, Milgrom2010, Milgrom2012, Bekenstein2004}, who proposed to modify the Newton gravity law for very low accelerations (MOND theory). Below $a_0 \sim c H_0/6$, he substituted the standard Newton gravity acceleration $g_N$ by $\sqrt{a_0 g_N}$. This causes that the Newton's law keeps valid for planetary and other small-scale systems but it does not apply to galaxies and galaxy clusters. As a consequence, rotation curves fit well observations \citep{Begeman1991}. Except for the flat rotation curves, the MOND theory is successful in accounting for some other phenomena that are difficult to explain using the DM hypothesis \citep{Sanders_McGaugh2002, Bugg2015}:

\begin{itemize}
\item
{\it{Faint satellite galaxies.}} The existence of the non-baryonic DM is questioned by a detailed study of properties of faint satellite galaxies of the Milky Way (MW), see Kroupa et al. \citep{Kroupa2010}, which are distributed on a planar structure. Similar alignments were observed also in isolated dwarf galaxies in the Local group \citep{Pawlowski_Kroupa_2013, Pawlowski_McGaugh_2014} as well as in more distant galaxies \citep{Galianni2010, Duc2014}. This is a challenge for cosmological simulations, because the DM sub-haloes are assumed to be isotropically distributed. 

\item
{\it{Dual dwarf galaxy theorem.}} The standard $\Lambda$CDM model predicts two types of dwarf galaxies: primordial DM dominated dwarf galaxies (Type A), and tidal and ram-pressure dwarf galaxies (Type B). While Type A dwarfs should surround the host galaxy spherically, the B dwarfs should be typically correlated in phase space. However, only dwarf galaxies of Type B are observed. This falsifies the Dual Dwarf Galaxy Theorem and the presence of DM haloes \citep{Kroupa2012}. 

\item
{\it{Baryonic Tully-Fisher relation.}} Observation of the baryonic Tully-Fisher (BTF) relation, which is a power-law relation between the rotation velocity of a galaxy and its baryonic mass $M_\mathrm{S+G}$, calculated as the sum of the stellar mass $M_\mathrm{S}$ and gas mass $M_\mathrm{G}$ \citep{Verheijen2001, Noordermeer_Verheijen2007, Zaritsky2014}. This empirical relation is valid over several orders of magnitude and with extremely small scatter. In the $\Lambda$CDM model, the rotation velocity should be primarily related to the total virial mass, represented mostly by the dark matter halo, $M \sim V^3$, but not to $M_\mathrm{S+G}$. Since the dark matter halo is largely independent of baryonic processes, it is difficult to explain the observed extremely low scatter of the BTF relation \citep{Lelli2016b, McGaugh2018, Lelli2019}. If the most mass of galaxy is formed by the baryonic dark matter located in the galaxy disk but not in the halo, the close relation between the stellar, gas and dust masses is expected.

\item
{\it{Radial acceleration relation.}} A further close connection between mass of stars and gas $M_\mathrm{S+G}$ and the total mass $M$ of galaxies, was revealed by McGaugh et al. \citep{McGaugh2016_PRL}, when they studied a relation between the acceleration $g_\mathrm{S+G}$ due to mass $M_\mathrm{S+G}$ and the observed acceleration $g_\mathrm{obs}$ due to total mass $M$. The observed relation is fully empirical and points to a strong coupling between the mass of the dark matter and mass of stars and gas. Similarly as for the BTF relation, the observed coupling between $g_\mathrm{S+G}$ and $g_\mathrm{obs}$ is difficult to explain in the $\Lambda$CDM model.

\end{itemize}

However, the alternative theories modifying either the Newton or Einstein gravity \citep{Mannheim2006} are used rather rarely being criticised for violating the Newton-Einstein gravitation theory. Although, the MOND theory matches some observations quite successfully, the theory was designed rather empirically to fit observations with no profound physical consistency. For example, some deep reasoning why value of $a_0$ is within an order of magnitude of $c H_0$ is missing. Note that the theory presented in this paper explains satisfactorily all the above mentioned phenomena similarly as MOND but with no need to violate the GR theory.

\subsection{Morphology of spiral galaxies}

Several theories have been proposed to explain the origin and evolution of structure of spiral galaxies and to predict basic properties of spiral arms \citep{Toomre1977, Dobbs_Baba_2014, Shu2016}. Lindblad \citep{Lindblad1962} was the first who assumed that the spiral structure arises from interaction between the orbits and gravitational forces of the stars of the disk and suggested to explain the spiral arms as density waves. This idea was further elaborated by Lin and Shu \citep{Lin_Shu_1964,Lin_Shu_1966} in their hypothesis of the quasi-stationary density waves, in which the spirals are formed by standing waves in the disk. They assumed that the spiral pattern rotates in a particular angular frequency different from the rotation velocity of stars, which depends on the star distance from the galaxy centre. The formation of the global spiral pattern is considered as an instability of the stellar disk caused by the self-gravity. The density-wave theory was further developed and extended \citep{Shu1970, Roberts1975, Sellwood_Carlberg_1984, Elmegreen1999, Sellwood2011} and it is now the main tool for studying the gravitational stability of disk galaxies. The results are not decisive but $N$-body simulations suggest that the spiral arms are transient and recurrent rather than quasi-stationary \citep{Baba2009, Baba2013}. 

The density-wave theory faces, however, with several open questions and limitations. First, the theory is based on the classical Newtonian gravity, which neglects the space expansion. Second, predictions of the theory are uncertain. Still it is not clearly resolved, whether the spiral arms must be dynamic or whether the quasi-static arms are a feasible solution. Third, the observationally documented growth of the spiral galaxies with cosmic time is completely ignored and unexplained in this theory. 

All the mentioned difficulties with modelling of spiral arms arise from the fact that the theory starts with rejecting an idea of spirals formed by stars and gas that remain fixed in the spirals. The reason is the so-called winding problem, when objects moving with the same orbital speed in the disk cause a differential rotation of material in galaxies. Since the length of orbits is shorter near the galaxy centre, the inner part of spirals winds up tighter than its outer part. Hence, a typical spiral pattern disappears after a few rotations. This idea is, however, simplistic and incorrect, because the GR effects in the galaxy evolution are ignored. If the space expansion and time dilation are considered, the galaxy size is growing and galaxy rotation speed varies with time. The spiral pattern is not destroyed, because it is continuously expanding. Consequently, the winding problem does not occur as demonstrated by numerical modelling in Section 3.

\subsection{Solar system}

Observations confirm that the global expansion affects also the solar system. Here we mention some of prominent examples \citep{Anderson1998, Iorio2015, Krizek2012, Krizek_Somer_2015, Krizek_Antigravity_2015}.

\subsubsection{Pioneer anomaly}

The Pioneer 10 and Pioneer 11 spacecraft were launched on 2 March 1972 and 4 December 1973 to explore the outer solar system \citep{Anderson1998}. Radiometric data received from the Pioneer 10/11 missions from distances of 20-70 AU indicated an anomalous blue-shifted frequency drift with a rate of $\sim 6 \times 10^{-9} \, \mathrm{Hz/s}$. The drift was interpreted as produced by an anomalous acceleration of the Pioneer satellites of $a = 8.74 \pm 1.33 \times 10^{-10} \, \mathrm{m s^{-2}}$. The acceleration is very close to the product of the current value of the Hubble constant and the speed of light \citep{Anderson2002, Turyshev2005, Turyshev_Toth_2010}: $H_0 c \approx 7 \times 10^{-10} \, \mathrm{m s^{-2}}$ .  This apparent violation of the Newton's gravitational law is known as the Pioneer anomaly and might originate in the global expansion of the Universe.

\subsubsection{Faint young Sun paradox}

According to the Standard solar model \citep{Bahcall2001}, the radius and luminosity of the Sun significantly evolved during the cosmic time. As the Sun is a star on the main sequence of the HR diagram, the solar radius was 4 Gyr ago about 89\% of the solar radius at present and the luminosity was about 73.8\% of the present luminosity (see Bahcall et al.\citep{Bahcall2001}, their tables 1 and 2). For a constant distance between the Sun and the Earth during this time span, such changes would have dramatic consequences for life conditions on the Earth \citep{Ribas2010}. The solar constant, i.e. the flux density at the Earth's mean orbital distance, is $I = 1.36 \, \mathrm{W m^{-2}}$ at present \citep{Kopp_Lean_2011}, but 4 Gyr ago it was $I_0 = 1.00 \, \mathrm{W m^{-2}}$ only. Calculating the equilibrium temperature as 
\begin{equation}\label{eq35}
T^{\mathrm{eq}} = \left(\frac{I(1-A)}{4\sigma}\right)^{1/4} \,, 
\end{equation}
where $A = 0.3$ is the Earth's albedo \citep{Goode2001} and $\sigma$ is the Stefan-Boltzmann constant, we get $T^{\mathrm{eq}} = 254.5 \, \mathrm{K}$ and $T_0^{\mathrm{eq}} = 236.0 \, \mathrm{K}$ for the present time and for the past, respectively. Assuming the same level of the greenhouse effect ($+32.5^\circ$C), the global average Earth's temperature would be $-4.7^\circ $C in the past instead of $13.9^\circ$C observed at present (https://www.climate.gov/news-features/understanding-climate/climate-change-global-temperature). In fact, because of the ice albedo, the temperature would be even lower in the era of 4 Gyr ago. For the Earth's albedo of 0.5, we get $T_0^\mathrm{eq} = 249.4 \, \mathrm{K}$ and the global average Earth's temperature would be $-23.8^\circ$C. By contrast, no glaciation is indicated from geological observations in the first 2.7 Gyr of the Earth's evolution \citep{Bertotti2003} and water-related sediments have been found 3.8 Gyr ago \citep{Windley1984}. This severe discrepancy is known as the Faint young Sun paradox.

The paradox is resolved, if the expansion of the solar system is taken into account \citep{Krizek2012, Krizek_Somer_2015}. The age of 4 Gyr corresponds to redshift $z = 0.46$ (see Fig. 1b) and the orbital radius of the Earth was $(1+z)$ times shorter than at the present time. The flux corrected for a shorter orbital radius is easily calculated as $1.00 \times (1+z)^2 = 2.13 \, \mathrm{W m^{-2}}$ and the corresponding Earth's temperature is $44.5^\circ \, \mathrm{C}$, provided we assume the same greenhouse effect as at present ($+32.5^\circ \, \mathrm{C}$). Hence, the temperature conditions on the Earth were convenient for life over the whole Earth's history. Note that higher temperatures of oceans ($70^\circ\mathrm{C}$) in the Precambrian era (3.5 Gyr ago) are independently indicated by observations of silicon and oxygen isotope data \citep{Knauth2005, Robert_Chaussidon_2006}. The recession velocity of the Earth from the Sun comparable to the Hubble flow is indicated also from growth patterns on fossil corals observed for the time span of the last 500 Myr \citep{Zhang2010}.

\subsubsection{Lunar orbit anomaly}

The Moon's orbital distance is slowly increasing and the Earth's rotation rate is decreasing due to tidal forces transferring angular momentum from the Earth to the Moon. In order to investigate the Earth-Moon system, the Lunar Laser Ranging Experiment (LLR) from Apollo 11, 14, 15 and Lunokhod missions was performed to measure accurately the recession velocity of the Moon. The missions report the Moon's semimajor axis $d = 384,402 \, \mathrm{km}$, which increases at rate \citep{Dickey1994} of $(3.82 \pm 0.08) \, \mathrm{cm/yr}$ . This value is anomalously high and inconsistent with an expected lunar recession velocity due to tidal forces, which should be lower  by $30 - 45\%$ \citep{Krizek2009, Riofrio2012, Krizek_Somer_2022b}. The observed lunar recession velocity would correspond to increasing Earth's rotation period at a rate of $+2.3 \, \mathrm{m s}$ per century, but only a rate of $+1.8 \, \mathrm{m s}$ per century is observed \citep{Stephenson2016}. In addition, numerical modelling of the orbital evolution under such tidal dissipation would imply the age of the lunar orbit to be $1.5 \times 10^9$ years, instead of $4 \times 10^9$ years suggested by observations \citep{Bills_Ray_1999}. This discrepancy is known as the Lunar orbit anomaly and so far its origin is unclear.  

If the expansion of the solar system is considered, the recession velocity of the Moon due to the expansion is $2.74 \, \mathrm{cm/yr}$ assuming distance of the Moon $d = 384,402 \, \mathrm{km}$ and the Hubble parameter $H_0 = 69.8 \, \mathrm{km s^{-1} Mpc^{-1}}$ (see, Freedman et al.\citep{Freedman2019}). Hence, the lunar recession velocity due to tides is reduced to $1.08 \, \mathrm{cm/yr}$ that is more realistic. If the tides were fully responsible for the slowing Earth's rotation at rate of $+1.8 \, \mathrm{m s}$ per century, the lunar recession velocity would be $\sim2.25 \, \mathrm{cm/yr}$. However, this is an upper limit only, because also other processes can slow down the Earth's rotation, such as impacts of massive meteorites, large earthquakes, huge volcanic eruptions, and energy dissipation in the Earth's mantle and the outer Earth's core due to convection.

\subsubsection{Other observations}

The Hubble flow in the solar system is indicated by many other observations \citep{Krizek_Somer_2015}. Mars had to be much closer to the Sun in the past; it is dusty and icy at the present, but detailed images of the Martian surface reveal that it was formed by rivers in the period of $3-4 \, \mathrm{Gyr}$ ago \citep{Carr1995, Davis2016, Salese2020}. Measurements of the Titan's orbital expansion rate by the Cassini spacecraft during ten close encounters of the Moon between 2006 and 2016 revealed that Titan rapidly migrates away from Saturn \citep{Lainey2020}. The Titan-Saturn mean distance is $d = 1 221,870 \, \mathrm{km}$ and the Titan's recession velocity is $11.3 \, \mathrm{cm/yr}$. The corresponding recession velocity due to the Hubble flow is $ 8.7 \, \mathrm{cm/yr}$. If this velocity is subtracted, the anomaly disappears and the resultant rate of $2.6 \, \mathrm{cm/yr}$ produced by tidal forces becomes realistic. Also, the expansion of the solar system can explain the formation of Neptune and Kuiper belt, the existence of fast satellites of Mars, Jupiter, Uranus and Neptune that are below the stationary orbit, or the large orbital momentum of the Moon \citep{Krizek_Somer_2015}.

\section{Discussion and conclusions}

The presented theory and numerical modelling satisfactorily explain several severe tensions between the standard $\Lambda$CDM model and observations:

\begin{itemize}

\item 
The improved Newtonian equations derived for the conformal FLRW metric predict an increasing radius of gravitational orbits. The rate of growth with cosmic time is $(1+z)^{-1}$. This applies to all local systems including galaxy clusters, galaxies, and planetary systems. The gravitational orbits are stationary in the comoving coordinates but non-stationary in the proper coordinates.

\item
The existence of spirals in disk galaxies is a direct consequence of the space expansion and time dilation. The spirals are formed by the stellar mass and gas that remain fixed in them.  The stellar mass and gas are continuously outflowed from the bulge-bar region into the spiral region. The spirals are detached from the bulge-bar region due to the space expansion.

\item 
Since the orbital velocity of particles is conserved and the radius of orbits gradually increases during the space expansion, the spiral galaxies display flat rotation curves. 

\end{itemize}

The constant rotation velocity, the space expansion and time dilation are the primary factors forming the morphology of spirals. The time dilation is particularly important, because a slower rate of time in the past significantly helps to separate spirals from the bulge-bar domain. As shown in the numerical modelling, the morphology of spirals depends on the expansion history of the Universe, on the galaxy mass, galaxy age and size of the bulge. Obviously, the presented modelling is far from being complete. A detailed parametric study based on observations of various types of galaxies is necessary for drawing more specific conclusions about the galaxy dynamics.

The previous theoretical attempts to correctly explain the galaxy dynamics were unsuccessful for the following reasons: (1) The simplest attempts ignored the space expansion and assumed galaxies in the static universe. Consequently, the GR effects related to the expansion of the Universe were neglected and stars moved along stationary orbits not evolving in time. (2) Theories, which considered the space expansion using the GR theory, applied an incorrect cosmological model defined by the standard FLRW metric. This metric erroneously assumes that time is invariant of the space expansion. This assumption has fatal consequences for dynamics of local systems. An additional radial acceleration originating in the space expansion is included in the Newtonian equations for orbiting bodies, but the proper angular momentum $L = R V^\theta$ keeps constant. The constant $L$ causes that the effect of the Hubble flow on the orbiting bodies is eliminated and the radius of the orbits is effectively insensitive to the space expansion. 

By contrast, the proper angular momentum $L = R V^\theta$ in the Newtonian equations derived under the conformal FLRW metric is not constant but it increases with redshift. Consequently, the orbits are not stationary any more but their radius increases with cosmic time. Since the velocity of orbiting massive particles does not depend on the space expansion and the radius of orbits is continuously increasing, the rotation curves are essentially flat. Importantly, the rotation curves are flat without assuming non-baryonic dark matter haloes surrounding galaxies. No dark matter is needed for explaining all basic properties of the galactic dynamics. Applying the conformal FLRW metric to interpretations of the SNe Ia dimming reveals that also dark energy is unnecessary for getting fit with observations \citep{Vavrycuk_Frontiers_2022}. Hence, dark energy and dark matter are false and superfluous concepts originating in a wrong description of the space expansion, when the time dilation is ignored during the evolution of the Universe. Once a correct metric is applied, the cosmological model is consistent with observations with no need to introduce new unphysical concepts. A controversy of dark matter and dark energy concepts is evidenced also by many other observations \citep{Koyama2016, Weinberg2013, Bull2016, Ezquiaga_Zumalacarregui2017, Kroupa2012, Kroupa2015, Buchert2016, Bullock_Boylan-Kolchin2017}. 

In addition, the presented theory resolves several other puzzles and paradoxes in cosmology. It explains the origin of spirals in a completely different way than proposed by the density-wave hypothesis. The spirals are not an effect of standing waves in the disk as so far believed \citep{Toomre1977, Dobbs_Baba_2014, Shu2016}. Instead, they are objects formed by material remained fixed in spirals. Still, the winding problem is avoided. The theory also removes tensions related to the observed galaxy growth explained by major and/or minor mergers of galaxies. Since the hypothesis of mergers \citep{Naab2009, McLure2013a, Kormendy_Ho_2013, Conselice2014} is refuted by observations of no evolution of the galaxy mass \citep{Bundy2017, Kawinwanichakij2020}, the problem of a galaxy growth is so far unsolved. The presented theory also resolves challenges to the $\Lambda$CDM model such as the problem of faint satellite galaxies, the baryonic Tully-Fisher relation or the radial acceleration relation. Furthermore, numerous puzzles in the solar system are successfully explained such as the Pioneer anomaly, the Faint young Sun paradox, the lunar orbit anomaly, the presence of rivers on ancient Mars, the Titan recession velocity anomaly, formation of the Kuiper belt and others \citep{Krizek_Somer_2015, Krizek_Antigravity_2015, Dumin2015}.

\appendix
\section{Gravitational orbits in the expanding universe described by the standard FLRW metric}

The metric tensor $g^{\mu \nu}$ of a gravitational field produced by a point mass $M$ situated in space obeying the standard FLRW metric reads (Noerdlinger and Petrosian \citep{Noerdlinger_Petrosian_1971}, their eq. 11)
\begin{equation}\label{eqA1}
ds^2 = -c^2(1+2\alpha)dT^2 + a^2(T) \left((1-2\alpha)\frac{dr^2}{1-kr^2}+r^2d\Omega^2\right) 
\end{equation}
where $T$ is the proper time and
\begin{equation}\label{eqA2}
\alpha = -\frac{G M}{r c^2}\,, \, |\alpha| \ll 1\,,
\end{equation}
is the Newtonian gravitational potential normalized to $c^2$, and $G$ is the gravitational constant. Assuming a massive nonrelativistic particle ($v \ll c$) orbiting in the gravitational field in the plane defined by $\phi = 0$ and calculating the Christoffel symbols $\Gamma_{\alpha \beta}^{\mu}$ in the geodesic equation (9), we get the following approximate equations
\begin{equation}\label{eqA3}
\ddot{r} - r {\dot{\theta}}^2 - \frac{G M}{\alpha^2 r^3} + 2\frac{\dot{a}}{a} \dot{r} =0 \,,
\end{equation}
\begin{equation}\label{eqA4}
r\ddot{\theta} + 2\dot{r}\dot{\theta} + \frac{\dot{a}}{a}r\dot{\theta}  = 0
 \,, 
\end{equation}
where dots over quantities mean derivatives with respect to time $T$. Inserting the proper distance $R = a(T) r$ into Eq. (A3) we get
\begin{equation}\label{eqA5}
\ddot{R} = -\frac{G M}{R^2} + R {\dot{\theta}}^2 + \frac{\ddot{a}}{a} R\,.
\end{equation}
Similarly, Eq. (A4) can be rewritten as
\begin{equation}\label{eqA6}
\frac{d}{dT} \left(a^2 r^2 \dot{\theta} \right) = \frac{d}{dT} \left(R^2 \dot{\theta} \right) = \frac{d}{dT} L = 0  \,, 
\end{equation}
where $L = R V^\theta$ is the proper angular momentum, and $V^\theta$ is the proper tangential velocity. Consequently, we can write (Carrera and Giulini \citep{Carrera_Giulini_2010}, their eq. 12ab)
\begin{equation}\label{eqA7}
\ddot{R} = -\frac{G M}{R^2} + \frac{L^2}{R^3} + \frac{\ddot{a}}{a} R\,,
\end{equation}
\begin{equation}\label{eqA8}
L = \mathrm{const}\,.
\end{equation}
The equations are called the modified Newtonian equations and they differ from the standard Newtonian equations describing the Kepler orbits by term $\frac{\ddot{a}}{a} R$ in Eq. (A7) related to the space expansion. The analysis of the modified Newtonian equations applied to the galaxy dynamics shows that assuming a constant $L$, the expansion term $\frac{\ddot{a}}{a} R$ affects the orbits within galaxies negligibly \citep{Faraoni_Jacques_2007}.



\end{document}